\documentclass[aps,prl,twocolumn]{revtex4}

\usepackage{graphicx}% Include figure files
\usepackage[hypertex]{hyperref}
\def\beq{\begin{equation}}
\def\eeq{\end{equation}}
\def\bea{\begin{eqnarray}}
\def\eea{\end{eqnarray}}

\def\msun{M_{\odot}}
\usepackage{color}

\begin{document}

\title{Dark Energy, Black Hole Entropy, and the First Precision Measurement \\in Quantum Gravity}

\author{Niayesh Afshordi}\email{nafshordi@perimeterinstitute.ca}
\affiliation{Perimeter Institute
for Theoretical Physics, 31 Caroline St. N., Waterloo, ON, N2L 2Y5,Canada}
\affiliation{Department of Physics and Astronomy, University of Waterloo, 200 University Avenue West, Waterloo, ON, N2L 3G1, Canada }

\date{\today}
%\preprint{astro-ph/yymmnnn}

\begin{abstract}
The two apparently distinct phenomena of dark energy (or late-time cosmic acceleration) and quantum gravity dominate physics on extremely low, and extremely high energies, but do not seem to have any apparent empirical connection. Nevertheless, the two have a theoretical connection, through the {\it cosmological constant problem}. I argue that the finite temperature quantum gravitational corrections to black hole entropy yields a pressure for the gravitational vacuum (or gravitational aether). Assuming that the relative corrections are linear in horizon temperature (i.e. are suppressed by one power of Planck energy), the pressure is comparable to that of dark energy for astrophysical black holes. This implies that the observation of late-time cosmic acceleration may have provided us with the first precision measurement of quantum gravity, i.e. that of black hole entropy.
\end{abstract}

\maketitle

The discovery of the late-time acceleration of cosmic expansion at the turn of the
century \cite{Riess:1998cb,Perlmutter:1998np}, and its interpretation as being due to a mysterious dark
energy component, sent shock waves through the theoretical physics
community, and is arguably the most puzzling aspect of modern
cosmology.

What puts this problem at the heart of modern physics and cosmology is that the simplest form of dark energy, or a cosmological constant, is predicted in the standard model of particle physics, but with a value that is some sixty orders of magnitude larger than the observed dark energy density. This is known as the {\it cosmological constant problem} \cite{1989RvMP...61....1W}, which suggests a yet-unknown connection between the largest and smallest physical scales ever probed, and thus its resolution could revolutionize our understanding of fundamental physics. In this {\it letter}, I will argue that such UV-IR connection may indeed exist through horizons of black holes.

To see this though, we should first review black hole thermodynamics. A mysterious discovery of the 20th century was that the classical general relativistic black holes appear to have a thermodynamic entropy proportional to their horizon area \cite{PhysRevD.7.2333,Hawking:1976de}. In fact, one can write analogs of all the laws of thermodynamics for the evolution of black holes \cite{1973CMaPh..31..161B}. In particular, the first law for a general Kerr-Newman black hole takes the form:
\beq dm =  T_H dS + \Omega dJ + \Phi dQ, \label{1st}\eeq
where $m$ is the black hole mass (or ADM energy), while
\beq
T_H = {\kappa \over 2\pi }, S = \frac{A}{4} \label{area}
\eeq
are horizon temperature and entropy, which in turn depend on surface gravity, $\kappa$, and area $A$ of the black hole horizon. Moreover, $\Omega$, $J$, $\Phi$, and $Q$ are the black hole angular frequency, angular momentum, electrostatic potential, and charge respectively.
For simplicity, we will focus on Schwarzschild black holes, for which we have:
 \beq
 T_{H} = \frac{1}{8\pi m}, S = \frac{1}{16 \pi  T_H^2},
 \eeq
while all the other constants vanish. However, note that the results below are only expected to change by dimensionless factors of order unity, if we relax this assumption. Also, note that throughout this {\it letter}, unless mentioned otherwise,  we use Planck units: $\hbar=c=G=1$.

Interestingly, Jacobson has even argued that Einstein's equations can be derived from the first law of thermodynamics for horizon areas, suggesting that the full General Relativity (GR), and not just black holes, might be a thermodynamic description of a more fundamental theory \cite{Jacobson:1995ab}. More recently, Verlinde provided a less technical account of this result for Newtonian gravity \cite{Verlinde:2010hp}, which was the intellectual motivation for this {\it letter}.

One thermodynamic quantity that is notably missing from the first law (Eq. \ref{1st}) is pressure. The reason is that the asymptotic space-time of Kerr-Newman black holes is Minkowski, which implies zero pressure, if one uses Einstein's equations. However, as I argue below, there might be reasons to think that this may not be accurate in a UV-complete quantum gravitational framework.

An intriguing approach to quantum gravity, known as emergent gravity, postulates that rather than being a fundamental symmetry of nature,  Lorentz symmetry is an emergent phenomenon at low energies, and the fundamental theory does not have this symmetry \footnote{A similar phenomenon is propagation of sound waves in a square lattice. Even though the fundamental theory is not isotropic, the low energy sound waves propagate isotropically.}. Some examples of this construction are \cite{Horava:2009uw, Gu:2009jh}. Also see \cite{Jacobson:2005bg, Maccione:2009ju} for experimental/astrophysical bounds on implications of such theories in particle physics.

Breaking Lorentz symmetry introduces a preferred frame of reference for the laws of physics. A covariant description of emergent gravity would promote this preferred frame into a fluid, which acts as a modern-day version of gravitational aether \cite{Jacobson:2004ts}. For example, the Horava-Lifshitz construction of emergent gravity \cite{Horava:2009uw} reduces to GR plus an incompressible fluid at low energies \cite{Afshordi:2009tt}.

While (by construction) the aether should become irrelevant in classical GR, the leading quantum gravity corrections might lead to a finite aether pressure. To see this, let us consider the corrections to the area law for black hole entropy (Eq. \ref{area}). While the area law implies that the black hole degrees of freedom live on the surface of the horizon, one may imagine that this surface has an intrinsic thickness of the Planck length. Therefore, as a toy model, we may assume that the true entropy of a black hole is proportional to the volume of a shell at the black hole horizon with Planck thickness, $\Delta V$. Given that the horizon radius is $2m = (4\pi T_H)^{-1}$ for a Schwarzschild black hole, this implies:
\bea S_{\rm toy} \equiv \frac{\Delta V}{4}= \frac{\pi}{3}\left[\frac{1}{(4\pi T_H)^3} -\left(\frac{1}{4\pi T_{H}}-1\right)^3\right] \nonumber\\
= \frac{1}{16\pi T_H^2} \left[1-4 \pi T_H+ \frac{1}{3}(4\pi T_H)^2\right].\label{stoy}
\eea
However, note that the sign of the leading correction depends on how we assume the mean radius of the shell changes with $T_H$.

Inspired by this toy model, we will assume finite temperature corrections to black hole entropy of the form:
\beq S = \frac{1}{16\pi T_H^2} \left[1+\alpha T_H+ {\cal O}(T_H^2)\right]. \label{sqg}\eeq
We further assume similar corrections to the mass-temperature relationship:
\beq m = \frac{1}{8\pi T_H} \left[1+\beta T_H + {\cal O}(T_H^2)\right],\label{mqg}\eeq
although, as we see below, $\beta$ will drop out of the leading correction to pressure.

We are now almost ready to use the first law of thermodynamics:
\beq T_HdS = dm + pdV, \label{1st+p}\eeq
for a Schwarzschild black hole to compute the finite-temperature pressure, $p$, of the gravitational aether. However, we also need an expression for, $V$,  the 3-volume of the horizon. This was found in \cite{Kastor:2009wy}, where a generalized first law for spherical black holes which includes variations in the cosmological constant in an Anti-de Sitter spacetime was derived. For a nearly flat asymptotic spacetime, it simply becomes:
\beq V \equiv \frac{4\pi r^3_{\rm obs}}{3} - \int^{r_{\rm obs}}_{r>r_H} d^3x \sqrt{-g}, \eeq
 which is the missing volume of the 3-space out to $r_{\rm obs}$ due the black hole horizon. One feature of this definition is that it is invariant under the change of coordinates, as long as the space-time is approximately flat at $r \sim r_{\rm obs}$. Moreover, we simply recover the flat space 3-volume for a Schwarzschild black hole:
\beq V = \frac{4}{3} \pi (2m)^3 = \frac{1}{48\pi^2 T_H^3} \label{vol_sch}.\eeq

Now, plugging Eq's (\ref{sqg}),(\ref{mqg}), and (\ref{vol_sch}) into (\ref{1st+p}), we find:
\bea &&p = \alpha \pi T_H^3 + {\cal O} (T^4_H) \nonumber\\\simeq && -\alpha p_{\Lambda, obs} \left(m \over 58~ M_{\odot} \right)^{-3} + {\cal O} (T^4_H).\label{pressure}\eea
Here, we should note that in \cite{Kastor:2009wy}, black hole mass is identified with its enthalpy, and not its energy. However, since $pV = {\rm const.}$ to leading order in Eq. (\ref{pressure}), the energy and enthalpy only differ by a constant, which does not affect our application of the first law (\ref{1st+p}).

In the second line of Eq. (\ref{pressure}), we compare the aether pressure to the observed pressure of dark energy:
\beq
p_{\Lambda, obs} = -\rho_{\Lambda, obs} = - 1.86 \times 10^{-29} (\Omega_{\Lambda} h^2) ~{\rm g/cm^{3}},
\eeq
where
\beq
\Omega_{\Lambda} = 0.728 \pm 0.016, h= 0.704 \pm 0.014 \label{data}
 \eeq
 quantify the current observational constraints on dark energy density and Hubble constant \cite{Komatsu:2010fb}. It thus becomes apparent from Eq. (\ref{pressure}) why I shall next entertain a connection between formation of astrophysical black holes and the observed dark energy phenomenon.

 A possible connection between astrophysical black holes and dark energy was first pointed out in \cite{Afshordi:2008xu}. It was later shown how this relationship could emerge in the spacetime of a static spherical black hole, surrounded by an incompressible gravitational aether \cite{PrescodWeinstein:2009mp}. To derive this, a Trans-Planckian ansatz that limits the redshift at the Schwarzschild radius to the ratio of Planck to Hawking temperatures was assumed \footnote{ In fact, the result is identical to Eq. (\ref{pressure}), if one sets the Trans-Planckian parameter $\theta_p$ (Eq. 20 in \cite{PrescodWeinstein:2009mp}) to $-2\alpha^{-1}$.}.

What is novel about the current thermodynamic derivation is that the result is quite independent of the microphysical model. It simply follows from a dimensional argument for finite temperature (or quantum gravity) corrections to black hole entropy, Eq. (\ref{sqg}), and the first law of thermodynamics, Eq. (\ref{1st+p}). Of course, nothing precludes setting $\alpha =0$, which might indeed be the case in some consistent quantum gravity theories. However, the discovery of late-time acceleration of cosmic expansion provides strong evidence to the contrary, i.e.:
\beq  \alpha = (-3.42 \pm 0.16) \times 10^{-3} \left(\overline{m}_{BH}\over 10 ~\msun\right)^3, \label{alpha_cons}\eeq
using the current observational constraints on cosmic acceleration (Eq. \ref{data}), and assuming an ``average'' astrophysical black hole mass $\overline{m}_{BH}$ (see below).

To find Eq. (\ref{alpha_cons}), we assumed that aether has only pressure and no density, which is the sufficient (but not necessary) condition for incompressibility, as in low energy Horava-Lifshitz theory. Moreover, the energy density of aether is significantly constrained by cosmological observations \cite{Afshordi:2009tt}. Given that acceleration depends on $\rho+3p$, if aether pressure satisfies:
\beq
3p = \rho_{\Lambda, obs}+3p_{\Lambda, obs} = 2p_{\Lambda, obs},
\eeq
it will yield the same cosmic acceleration as a cosmological constant with energy density $\rho_{\Lambda, obs}$  \footnote{A more rigorous version of this statement is derived in \cite{PrescodWeinstein:2009mp}.}. We stress that unlike dark energy, by construction, aether has no energy density. However, due to the presence of black holes, the standard continuity equation in FRW spacetimes does not apply, and thus pressure can be (nearly) constant but non-vanishing.

The main obstacle in interpreting Eq. (\ref{alpha_cons}) as a cosmological measurement of black hole entropy is the relevant $\overline{m}_{BH}$, given that black holes of different masses exist in our universe. Since different black holes have different horizon temperatures, the system is clearly not in thermal equilibrium. Therefore, thermodynamic arguments will  not be sufficient to provide an answer. One way to approach this problem is to assume that the aether pressure is set by Eq. (\ref{pressure}) close to each black hole, and then it is in approximate hydrostatic equilibrium in between the black holes, assuming that sound crossing time is much shorter than the dynamical time of the system.   
Within this assumption, \cite{PrescodWeinstein:2009mp} shows that the mean aether pressure in a quasi-static space-time of multiple black holes depends on an effective mass $\overline{m}_{BH}$, where:
\beq
\ln \overline{m}_{BH}  \equiv \langle \ln m_{BH} \rangle_{\rm mass~weighted} = \frac{\sum_i m_{i} \ln m_{i}}{\sum_i m_{i}}.\label{mbar}
\eeq
In other words, $\overline{m}_{BH}$ is the mass-weighted geometric average of black hole masses. Since most of the mass of of astrophysical black holes are in stellar black holes, they dominate the contribution to $\overline{m}_{BH}$, implying that $\overline{m}_{BH} \sim 10 \msun$. However, late-time accretion of mass into supermassive black holes could yield $\overline{m}_{BH}$ as high as $25 \msun$ \cite{PrescodWeinstein:2009mp}. For this range, Eq. (\ref{alpha_cons}) implies:
\beq
 -0.05 \lesssim \alpha \lesssim -0.003.
\eeq 

 We should also point out that incompressible gravitational aether appears in a modification of Einstein's gravity that decouples vacuum density from geometry \cite{Afshordi:2008xu}, thus avoiding the old cosmological constant problem \cite{1989RvMP...61....1W}. However, solutions in this theory that include both ordinary matter and black holes are yet non-existent.

Taken at face value, and barring a sheer numerical coincidence, our finding provides a novel challenge for quantum gravity theories, which is fundamentally different from the traditional criteria of renormalizability or low energy tests. In particular, any description of gravity as a locally covariant effective action for metric is expected to behave like:
\beq
{\cal S}=\!\int d^4x\sqrt{-g}\left[\Lambda + \frac{R}{16\pi} + {\cal O}(R^2, R^{\mu\nu}R_{\mu\nu},R^{\mu\nu\alpha\beta}R_{\mu\nu\alpha\beta}) \right],
\eeq
i.e. the leading relative corrections to GR scale as curvature, $R^{\mu\nu\alpha\beta} \sim T^2_H$ in Planck units. This is consistent with the fact that relative corrections to the entropy in many local theories of quantum gravity scale as $A^{-1} \log A \sim -T^2_H \log T_H$ (e.g., see \cite{Modesto:2010rm}). Therefore, we expect $\alpha=0$ in local theories (at least at the perturbative level), which implies $p \sim T^4_H$ from Eq. (\ref{pressure}), and is consistent with Hawking radiation field.

However, note that the assumption of presence of aether already precludes local covariance. For example, one may expect corrections suppressed by one power of mean extrinsic curvature of the aether comoving hypersurfaces, $K$, which scales as $T_H$ on dimensional grounds. As an example, Let us consider a modification of the GR action of the form:
\beq
{\cal S}_f = {\cal S}_{GR} + \int d^4x \sqrt{-g} f(K), \label{fk_action}
\eeq
where
\beq
f(K) = \alpha' K^3 + {\cal O}(K^4),\label{fk}
\eeq
if we want the corrections to be analytic in $K$ and Planck suppressed. Generalizing the method outlined in \cite{Afshordi:2009tt}, we can write ${\cal S}_f $ as:
\bea
{\cal S}_f = {\cal S}_{GR} &+& \int d^4x \sqrt{-g} f(K)
= {\cal S}_{GR} - \int d^4x \sqrt{-g} \left[\varphi K +V(\varphi)\right] \nonumber\\
= {\cal S}_{GR} &+&\int d^4x \sqrt{-g} \left[\sqrt{\partial^\mu \varphi \partial_\mu \varphi}  -V(\varphi)\right],
\eea
i.e. a cuscuton action \cite{Afshordi:2006ad}, where $V(\varphi)$ is the Legendre transform of $f(K)$:
\beq
K=-\frac{\partial V}{\partial \varphi}, f(K) = -\varphi K - V(\varphi),
\eeq
while the preferred foliation in ${\cal S}_f $ coincides with constant $\varphi$ hypersurfaces for reasonable boundary conditions (see \cite{Afshordi:2009tt} for a discussion). In other words, the preferred foliation (or aether comoving) hypersurfaces in ${\cal S}_f $ are constant mean curvature (CMC) surfaces. Also, notice that constraint algebra closes as Einstein+cuscuton field equations can be solved consistently, at least prior to black hole formation \cite{Afshordi:2007yx}. 

For $f(K)$ in Eq. (\ref{fk}) one finds:
\beq
V(\varphi) = \pm \frac{2\varphi^{3/2}}{(27 \alpha')^{1/2}} + {\cal O}(\varphi^2/\alpha').
\eeq

There is an interesting corollary from the emergence of CMC hypersurfaces in ${\cal S}_f$ action. It was shown in \cite{2001PhRvD..63f4024G} that in the Schwarzschild spacetime, CMC surfaces of large negative $K$ can get arbitrarily close to the central singularity, while CMC surfaces of large positive $K$ barely penetrate the horizon, and pile up just inside of it \footnote{Note that \cite{2001PhRvD..63f4024G} uses an opposite sign in the definition of $K$.}. For large $K$, the CMC surfaces only penetrate the horizon by the proper distance (or time) of $K^{-1}$.
Therefore, the quantum corrections to the static classical spacetime (according to action ${\cal S}_f $) only become important close to the singularity, as well as within a shell of Planck thickness inside the Schwarzschild radius. Surprisingly, the latter brings us exactly back full circle, to the toy model of Eq. (\ref{stoy})! This is also a unique feature of foliation-violating theories of quantum gravity, as the spacetime remains classical at horizon in local quantum gravities, at least at the perturbative level. A notable exception might be the ``fuzzball'' proposal in string theory, where the spacetime is quantum mechanical everywhere inside the horizon as a result of (non-perturbative) tunneling effects (see e.g., \cite{Mathur:2008nj} for a review).

To conclude, I have argued that possible quantum corrections to black hole entropy that are suppressed by one power of Planck energy imply a novel quantum hair for the black hole, a pressure for gravitational vacuum/aether that scales as the cube of horizon temperature in Planck units. This is comparable to the observed dark energy pressure for astrophysical black holes, suggesting the exciting conclusion that observation of late-time cosmic acceleration might be {\it the first precision measurement in quantum gravity}. It also provides support for emergent or aether theories of gravity, as this type of correction is absent in local theories of quantum gravity, at the perturbative level.

The author is grateful to Latham Boyle,  Leonardo Modesto, Rob Myers, Lee Smolin, Rafael Sorkin, and Jennie Traschen for invaluable discussions and comments. Research at Perimeter Institute is
supported by the Government of Canada through Industry Canada and by the
Province of Ontario through the Ministry of Research \& Innovation.

\bibliography{precision_QG}
\end{document}